\documentclass[11pt]{article}
\pdfoutput=1

\usepackage{setspace}
\usepackage[letterpaper,total={15.5cm, 22.9cm}]{geometry}
\usepackage{titlesec}
\titleformat*{\section}{\Large\bfseries}
\titleformat*{\subsection}{\large\bfseries}
\titleformat*{\subsubsection}{\bfseries}
\usepackage[tikz]{mdframed}

\newcounter{question}
\stepcounter{question}

\mdfdefinestyle{questionstyle}{outerlinewidth = 0.2 , roundcorner = 2pt , leftmargin = 4 , rightmargin = 4 , backgroundcolor = gray!10 , linecolor = black!60 , innertopmargin = 5pt , innerbottommargin = 10pt , everyline = true , splittopskip = 18pt , splitbottomskip = 10pt}
\newcommand{\question}[1]{
\begin{mdframed}[style=questionstyle]
\quad #1
\end{mdframed}
\stepcounter{question}}

\usepackage{microtype}
\usepackage{upgreek}
\usepackage{empheq}
\usepackage{amsmath,amsthm,amssymb}
\usepackage{booktabs}
\usepackage{cite}
\usepackage{enumitem}

\usepackage{ifpdf}
\usepackage{lmodern}
\usepackage[T1]{fontenc}
\usepackage[ansinew]{inputenc}
\usepackage{relsize}
\usepackage{comment}
\usepackage{graphicx}
\usepackage{color}
\definecolor{darkblue}{cmyk}{0.9,0.9,0,0}
\definecolor{darkgreen}{rgb}{0,0.55,0}
\usepackage[colorlinks=true,linkcolor=darkblue,citecolor=darkblue,urlcolor=darkblue]{hyperref}
\usepackage{epsfig}
\usepackage{epstopdf}
\usepackage{mathtools}

\makeatletter
\long\def\@makecaption#1#2{
  \vskip\abovecaptionskip
  \sbox\@tempboxa{{\captionfonts #1: #2}}
  \ifdim \wd\@tempboxa >\hsize
    {\captionfonts #1: #2\par}
  \else
    \hbox to\hsize{\hfil\box\@tempboxa\hfil}
  \fi
  \vskip\belowcaptionskip}
\makeatother

\usepackage{array, booktabs, makecell, multirow}
\setcellgapes{3pt}
\def\ni{\noindent}

\def\cN{{\cal N}}

\def\b{\beta}
\def\g{\gamma}

\def\l{\lambda}

\def\BH{{\rm BH}}
\def\max{{\rm max}}
\def\S{\mathbb{S}}
\def\N{\mathbb{N}}

\def\x{\times}

\def\1{{\rm 1-loop}}

\def\c{\cite}

\def\cN{\mathcal{N}}

\def\c{\cite}

\def\vs{\vskip .1 in}

\def\o{\over}
\def\g{\gamma}
\def\D{\Delta}
\def\rar{\rightarrow}

\def\eqr{\eqref}

\def\O{{\cal O}}

\def\ssec{\subsection}
\def\sssec{\subsubsection}
\def\sec{\section}
\def\i{\infty}
\def\foot{\footnote}
\newcommand{\es}[2] {\begin{equation} \label{#1} \begin{split} #2 \end{split} \end{equation}}
\newcommand{\e}[2] {\begin{equation} \label{#1} #2 \end{equation}}

\newcommand{\beq}{\begin{equation}}
\newcommand{\eeq}{\end{equation}}
\newcommand{\beqy} {\begin{eqnarray}}
\newcommand{\eeqy} {\end{eqnarray}}
\newcommand{\bsmat}{\begin{smallmatrix}}
\newcommand{\esmat}{\end{smallmatrix}}
\newcommand{\bmat}{\begin{matrix}}
\newcommand{\emat}{\end{matrix}}

\def\({\left(}
\def\){\right)}
\def\[{\left[}
\def\]{\right]}

\def\<{\langle}
\def\>{\rangle}

\def\b{\beta}
\def\g{\gamma}

\def\d{\delta}
\def\D{\Delta}

\def\l{\lambda}

\def\vs{\vskip .1 in}

\usepackage{varioref}
\usepackage{makeidx}
\makeindex

\usepackage[english]{babel}
\usepackage{subfig}

\numberwithin{equation}{section}

\usepackage{tabularx}

\begin{document}

\begin{spacing}{1.15}

\begin{titlepage}

\vspace*{3cm}
\begin{center}
{\LARGE \bfseries A Rigorous Holographic Bound on \\ \vs AdS Scale Separation}

\vspace*{1cm}

Eric Perlmutter

\vspace*{2mm}

\textit{\small Universit\'e Paris-Saclay, CNRS, CEA, Institut de Physique Th\'eorique, 91191, Gif-sur-Yvette, France\\
\vskip .1 in
Institut des Hautes \'Etudes Scientifiques, 91440, Bures-sur-Yvette, France}

\vspace{2mm}

{\tt \small perl@ipht.fr}

\vspace*{1cm}
\end{center}
\begin{abstract}

We give an elementary proof of the following property of unitary, interacting four-dimensional $\mathcal{N}=2$ superconformal field theories: at large central charge $c$, there exist at least $\sqrt{c}$ single-trace, scalar superconformal primary operators with dimensions $\Delta \lesssim \sqrt{c}$ (suppressing multiplicative logarithmic corrections). This follows from a stronger, more refined bound on the spectral density in terms of the asymptotic growth rate of the central charge. The proof employs known results on the structure of Coulomb branch operators. Interpreted holographically, this bounds the possible degree of scale separation in semiclassical AdS$_5$ half-maximal supergravity. In particular, the bulk must contain an infinite tower of charged scalar states of energies parametrically below the large black hole threshold, $E_{\rm BH} \sim c$. We address the extreme case of AdS$_5$ pure supergravity, ruling it out as the asymptotic limit of certain sequences in theory space, though the general question remains open.

\end{abstract}

\end{titlepage}
\end{spacing}

\pagenumbering{roman}
\begin{spacing}{0.7}

\setcounter{tocdepth}{1}
\tableofcontents
\end{spacing}


\pagenumbering{arabic}
\setcounter{page}{1}

\begin{spacing}{1.11}

\sec{The Problem}

The question of AdS scale separation is whether a consistent theory of quantum gravity can, in the semiclassical limit, admit an AdS vacuum without parametrically large extra dimensions. 

Recent years have seen significant efforts to scrutinize and extend earlier proposals for scale-separated solutions of string theory \c{Kachru:2003aw,DeWolfe:2005uu,Balasubramanian:2005zx}. This uptick was spurred in part by the AdS distance conjecture \c{Lust:2019zwm} and other Swampland considerations, and the pressing cosmological implications of this question vis-\`a-vis possible de Sitter uplift.

A direct bulk approach to the problem is inherently limited: to tuck the extra dimensions away at a parametrically small scale, one must grapple with the full machinery of string/M-theory, not just its 10- or 11-dimensional supergravity limit. For this reason, it seems fair to say that gravity constructions are unlikely to definitively answer this question beyond a reasonable doubt, unless and until non-perturbative string/M-theory are well-understood.

Boundary CFT approaches, however, offer hope for unambiguous resolution. There is a growing appetite for the conformal bootstrap to solve this problem, using axiomatic constraints to rigorous ends. In particular, one might realistically hope to exclude, or bound the degree of, AdS scale separation holographically using the large $N$ bootstrap (as opposed to explicitly constructing a candidate CFT). Some older \c{Aharony:2008wz,deAlwis:2014wia} and newer \c{Conlon:2018vov,Conlon:2021cjk,Apers:2022zjx,Apers:2022tfm,Apers:2022vfp} works explore the holographic dictionary for putative scale-separated string backgrounds, but fundamental constraints are missing. A collage of constructions, claims, counter-claims and conjectures is well-reviewed in \c{Coudarchet:2023mfs}.

For better or worse, the conformal bootstrap is not magic: the CFT avatar of excluding AdS scale separation is still a hard problem. As emphasized in \c{PS,AP}, this question differs fundamentally from the canonical bootstrap endeavor of maximizing the spectral gap to the first primary operator: large (i.e. AdS-sized) extra dimensions in gravity are manifest as {\it infinite towers} of operators in CFT. Their characteristic energy $\D$ and asymptotic density $\rho(\D)$ encode the size and number of bulk dimensions, respectively, quantities which can be read off from a positive sum rule for CFT correlators of light operators \c{AP}. 

Even supersymmetric versions of this problem, as articulated in \c{Gopakumar:2022kof}, remain open, providing compelling targets for the superconformal bootstrap. It is not known whether continuous R-symmetry of a superconformal field theory (SCFT) is always geometrized in AdS, despite familiar examples under the lamppost. Does quantum gravity admit, say, a maximally supersymmetric ``pure AdS$_5$'' background without a large $S^5$? There are arguments both for \c{ac} and against \c{mrv} this possibility in the literature. The analogous questions on the CFT side concern the existence of ``exotic'' SCFTs without infinite towers of BPS operators, which are a boundary hallmark of a large compact manifold in the bulk. 

This paper takes a step forward on the supersymmetric version of this problem, proving a universal result for 4d $\cN=2$ SCFTs at large central charges. Our proof uses nothing more than established facts about Coulomb branch geometry. The main result is in \eqr{zbound}--\eqr{absbound}, a constraint on the spectrum of Coulomb branch primaries. Extremizing over  theory space gives the absolute bound \eqr{absbound}, valid for all $\cN=2$ SCFTs (subject to widely-held assumptions detailed below), as quoted in the Abstract in slightly simplified form. 

The result is modest, and its proof, elementary; but its consequences for gravity are significant. By AdS/CFT, these theories are dual to asymptotically AdS$_5$ vacua of semiclassical gravity with (at least) half-maximal supersymmetry. The result establishes the necessity of an infinite (Planckian) number of $U(1)$-charged, scalar bulk states which are parametrically below the large AdS$_5$ black hole threshold, $E_\BH \sim c$. For $a\approx c$, this is a quantitative bound on the degree of scale separation in AdS$_5$ Einstein supergravity. We address the extreme sub-case of a putative theory of semiclassical AdS$_5$ pure supergravity, which we define (after some discussion in Appendix \ref{app}) as a theory with only supergravitons parametrically below the five-dimensional Planck scale: though we cannot unequivocally rule this spectrum out, we do exclude it as the asymptotic limit of certain paths through superconformal theory space. 

\sec{The Ingredients}

We consider unitary, local $\cN=2$ SCFTs in four dimensions; this is a vast subject (see e.g. \c{beemrev,hitch}), but we will keep our presentation to the point. We are interested in the BPS spectra of these theories at large central charge. An $\cN=2$ SCFT is partially characterized by its Coulomb branch, of rank $r\in\mathbb{N}$. The Coulomb branch may be parameterized by the vacuum expectation values of $r$ superconformal primaries, the generators of the chiral ring, which are the bottom components of protected $\mathcal{E}_{r(0,0)}$-type superconformal multiplets in the notation of \c{Dolan:2002zh} ($L\overline{B_1}[0;0]^{(0;r)}$ multiplets in the notation of \c{Cordova:2016emh}). These operators, call them $\{\O_i\}$, have vanishing $SU(2)_R$ charge, and $U(1)_r$ charge equal to their conformal dimensions, $r_i = \D_i$. In all known examples, the $\{\O_i\}$ are Lorentz scalars, which we have assumed here.

Let us state up front that we are making the oft-used assumption that the Coulomb branch is freely generated, i.e. that the generators satisfy no non-trivial relations.\foot{We thank Mario Martone and Leonardo Rastelli for discussions about the status of this assumption.} This allows for an unambiguous assignment of dimensions $\D_i$ to operators $\O_i$. There is a whole story here, with possible violations of this property if discrete gaugings are allowed, leading to complex singularities of the Coulomb branch geometry \c{Argyres:2017tmj,Argyres:2018wxu,Bourget:2018ond,Argyres:2020nrr}. Having emphasized this, we henceforth take the Coulomb branch to be freely generated, while keeping in mind the limitations of this assumption at the most abstract level of $\cN=2$ SCFT.

We will need two ingredients for our proof.

\sssec*{\bf i) Central charge sum rule} 

Recent work on the intricacies of Coulomb branch geometry has established the following central charge formulae in $\cN=2$ SCFTs \c{mario}:
\es{ccform}{2a &= -{r\o 12}  + {\S(r) \o 2} + f(\text{\tt other Coulomb branch data})\\
c &= {r\o 6} +  f(\text{\tt other Coulomb branch data})}
where
\e{}{\S(r) := \sum_{i=1}^r \D_i}
Happily, we won't need the unspecified function of {\tt other Coulomb branch data} in order to prove the desired result at large central charge. Instead we take a simpler approach. First, these results imply the following formula, previously established in a more limited context by Shapere and Tachikawa \c{ST}:
\e{st}{2a-c = { \S(r)\o 2}-{r\o4}\,,\quad \text{where}\quad \S(r) := \sum_{i=1}^r \D_i}
Noting the unitarity bound $\D_i \geq 1$, saturated by a free vector multiplet, an interacting unitary $\cN=2$ SCFT obeys $\S(r) > r$ and hence $2a-c > {r\o 4}$. Besides finding the independent formulae \eqr{ccform}, one achievement of the work since \c{ST} is that it establishes the Shapere-Tachikawa formula \eqr{st} in fuller generality, without the assumptions of \c{ST} (e.g. that there is a Lagrangian point on the SCFT moduli space).\foot{The work \c{mario} and its predecessors make their own assumptions, of course, one of which (free generation) was discussed earlier. Nevertheless their results are expected to be very robust, sufficient to perform a quite general classification of $\cN=2$ SCFTs. In indirect support of this, the rationality of central charges was confirmed by \c{Rastelli:2023sfk} using different, Higgs branch-based methods and employing independent assumptions.} We will use this formula in conjunction with the Hofman-Maldacena bound \c{Hofman:2008ar} for local $\cN=2$ SCFTs,
\e{hm}{{1\o2} \leq {a\o c} \leq {5\o 4}\,.}
The lower and upper bounds are realized by free hypermultiplets and free vector multiplets, repectively. 

\sssec*{{\bf ii) Coulomb branch operator dimensions}}

Our second ingredient is a fascinating feature of $\cN=2$ SCFT operator spectra: at a given rank $r$, the Coulomb branch dimensions $\{\D_i\}$ are drawn from a finite, $r$-dependent set of rationals. As proven in \c{Argyres:2018urp,Caorsi:2018zsq}, this set admits a remarkably concise characterization:
\e{}{\D_i \in \left\{ {n\o m}\,\Big |\, \varphi(n) \leq 2r\,,~~ 0<m\leq n\,,~~ (m,n)=1\right\}}
where $\varphi(n)$ is Euler's totient function. Most of the number-theoretic structure of this result will not be used here. All we need is the existence of $\D_\max$, the largest possible dimension at rank $r$: 
\e{}{\D_\max := \text{max}\,\big\{n\,|\, \varphi(n) \leq 2r\big\}}
For finite $r$, $\D_\max$ is finite. In what follows, we order the dimensions in a given $r$-tuple as 
\e{}{1\leq \D_1 \leq \D_2 \leq \cdots \leq \D_r \leq \D_\max\,.}

\ssec{Large $r,a,c$ limits}

Consider the large rank limit, $r\rar\i$. We first establish the following warmup result:\foot{Throughout the paper we use $\sim$ to indicate asymptotic scaling up to numerical prefactors, and $\approx$ to indicate asymptotic equality including numerical prefactors, with additive corrections left implicit in all cases.} 
\vs

\centerline{\it $\S(r) \sim r$ is inconsistent with the existence of the SCFT at large rank.}

\ni More precisely, $\S(r) \sim r$ is incompatible with a finite partition function on the spatial sphere. The proof is simply that a unitary theory can only achieve linear scaling of $\S(r)$ if $\O(r)$ of the $r$ Coulomb branch primaries have dimensions bounded above by $\D_* \sim \O(1)$; but this violates finiteness of $Z_{S^1_\b \x S^3} = {\rm tr}_{\mathcal{H}}(e^{-\beta H})$, which requires that
\e{linear}{\int_0^{{\rm finite}} d\D\, \rho(\D) < \i}
where $\rho(\D)$ is the spectral density of local operators on $S^3$. Therefore, the large $r$ limit does not exist.\foot{Let us register a complementary (but more restricted) argument that $a \approx c \sim r$ theories cannot be dual to semiclassical Einstein gravity specifically:  with this scaling, the species scale $\Lambda_s = (c/n_L)^{1/3}$, where $n_L$ is the number of bulk fields with energies below $\Lambda_s$, would become $\O(1)$, invalidating Einstein gravity as a good semiclassical EFT.} One can rephrase this conclusion as the absence of ``large $c$ $\cN=2$ vector models''; we suggest that this may also be true of $\cN=1$ SCFTs, for a suitable replacement of the Coulomb branch rank.\foot{The works \c{Agarwal:2019crm,Agarwal:2020pol} study putative 4d $\cN=1$ SCFTs, labeled ``dense'' theories, which have $a \sim c \sim N$, where $N$ is the rank of a UV gauge theory from which the IR SCFT flows. These theories have bands of operators within finite $\O(1)$ windows with $1/N$-suppressed splittings, which therefore become continuous in the large $N$ limit, violating the finiteness criterion \eqr{linear}.} 

So, at $r\rar\i$ we have $\S(r) \gg r$, and hence
\e{eq29}{2a-c \approx {\S(r)\o 2}}
Therefore, $r\rar\i$ implies $2a-c\rar\i$. 
Combining this with \eqr{hm} implies that $a$ and $c$ both grow large independently:
\e{}{r\rar\i \qquad \Rightarrow \qquad a,c \rar\i}
That is, {\it large rank implies large central charges}. 

The implication runs in reverse as well, modulo two small caveats. The first is if $2a-c$ is finite to leading order: in particular, $r$ can remain finite at large central charge if and only if
\e{ac12}{{a\o c} \approx {1\o 2} + \O\({1\o c}\)}
as $c\rar\i$ (with a strictly positive correction term). The case ${a\o c} = {1\o 2}$ is only known to be realized by free hypers; adopting this perspective, the possibility \eqr{ac12} can be phrased as the theory ``becoming free'' at large central charge. In addition, one must note the logical possibility that the SCFT at large central charge could in principle have no Coulomb branch, i.e. $r=0$; but it is widely believed that there are no interacting $\cN=2$ SCFTs with $r=0$.

All told, we can summarize as follows: assuming the existence of a Coulomb branch,
\e{}{a,c\rar\i \qquad \Leftrightarrow \qquad r\rar\i\,,}
with the only possible exception to the $\Rightarrow$ direction being the edge case ${a\o c} \approx {1\o 2}$. This case is, at any rate, maximally far from the Einstein regime in which $a \approx c$, to which we soon turn. 

As for the Coulomb branch dimensions at $r\rar\i$, the upper bound $\D_\max$ scales asymptotically as \c{Argyres:2018urp}
\e{dmaxr}{\D_\max \approx 2e^{\g_E} \,r \log\log r}
where $\g_E\approx 0.577$ is the Euler gamma function. 

\sec{The Proof} 

The key observation is that the upper-boundedness of Coulomb branch dimensions, which in turn determine the central charges, implies a spectral bound of the former when expressed in terms of the latter. 

In particular, bounding spectra at large central charge becomes an extremization problem to be solved in the asymptotic limit $r\rar\i$: {\it given an ordered $r$-tuple $\{\D_1,\ldots,\D_r\}$, maximize this set as a function of central charges, where the latter are determined by $\S(r)$ via \eqr{st} and \eqr{hm}.} We are using ``maximize this set'' as a placeholder for different choices of extremization, depending on the problem of interest. For instance, to extremize the spectral gap  \`a la bootstrap, one solves a {\it maximin problem}, choosing to maximize $\D_1$ among allowed $r$-tuples.

The main result now follows quickly. We established earlier that $\S(r) \gg r$, but we now note the obvious upper bound as well, 
\e{}{\S(r) \leq r \D_\max \approx 2e^{\g_E} r^2 \log\log r\,.}
In fact, strictly maximizing all $r$ operator dimensions is disallowed by more subtle number-theoretic aspects of Coulomb branch geometry -- when $\D_i = \overline \D$ for some $\overline \D$, the latter must actually be drawn from the allowed set at $r=1$ \c{Argyres:2018urp} -- so in the $r\rar\i$ limit, one ought to interpret the upper bound as 
\e{smax}{\S(r) \approx r \D_\max}
with negative splittings that are subleading in $r$. For simplicity, and our ultimate interest in the Einstein gravity regime, we specialize to $a\approx c$. Then plugging \eqr{dmaxr} and \eqr{smax} into \eqr{eq29} gives
\e{c2loglog}{c \approx e^{\g_E} r^2 \log\log r\,.}
In terms of central charge,
\e{}{\D_\max \approx 2e^{\g_E/2} \sqrt{c} \sqrt{\log\log c}\,.}
Therefore, {\it there are $r \approx e^{-\g_E/2} {\sqrt{c}/ \sqrt{\log\log c}}$ operators with dimensions $\D_i \leq \D_\max$.} Simplifying this exact statement by dropping the logs and prefactors gives the result quoted in the Abstract. This clearly generalizes away from $a\approx c$ to any asymptotic ratio of $a/c$ at large rank: thanks to Hofman-Maldacena, $a$ and $c$ will have identical $r$-scaling, but different numerical prefactors determined by their $\O(1)$ ratio. 

More generally, we can classify theories by the growth of central charge with the rank:
\e{zscal}{\S(r) \sim r^z\,,\quad 1 \leq z \leq 2\,.}
Multiplicative logarithmic enhancements -- required at $z=1$ (cf. \eqr{linear}) and possible at $z>1$ (e.g. cf. \eqr{c2loglog}) -- are left implicit to avoid clutter. Then for any finite $2a-c>0$, whereupon $a\sim c\sim r^z$, the extremal spectrum consistent with this scaling, i.e. the sparsest possible low-lying spectrum, is obtained by taking the lowest $\approx r$ operator dimensions to all approach 
\e{}{\D_*(z) \sim r^{z-1}\,.}
This corresponds to parametric solution of a constrained optimization problem: maximize $\D_r$, then maximize $\D_{r-1}$, and so on until maximization of $\D_1$, all subject to the scaling \eqr{zscal}. In other words, defining a spectral function $\N(\D_-,\D_+)$ counting primaries in an interval,
\e{}{\N(\D_-,\D_+) := \int_{\D_-}^{\D_+} d\D \,\rho(\D)\,,\quad \N_*(z) := \N(1,\D_*(z))}
where $\rho(\D)$ is the spectral density of Coulomb branch generators, the scaling \eqr{zscal} requires that $\N_*(z) \approx r$, and the extremal spectrum has these states ``piled up'' near $\D_*(z)$.\foot{Equivalently, $\N(\D_-,\D_\max) \sim o(r)$ for any $\D_- \gg \D_*(z)$: there can be states with $\D \gg \D_*(z)$, but not too many of them, lest the scaling \eqr{zscal} be violated.} Therefore, and restating in terms of central charge $c\rar\i$, 
\question{
\e{zbound}{c\sim r^z \quad \Rightarrow \quad \exists~~ \N_*(z)~\text{operators}~ \{\O_i\}~\text{with}~\D_i\lesssim  \D_*(z)}
where
\e{zboundscal}{\N_*(z) \sim c^{1/z}\,,\quad \D_*(z) \sim c^{1-{1/z}}}
}
\vs
\ni This is the general result. Scanning over possible SCFTs by varying $z$, the spectral gap as a function of $c$ is maximized at $z=2$ dressed with doubly logarithmic corrections \eqr{c2loglog}, giving the absolute bound derived earlier, valid for all $\cN=2$ SCFTs:
\question{\e{absbound}{\exists~~ \N_*\approx e^{-\g_E/2} {\sqrt{c}\o \sqrt{\log\log c}}~\text{operators}~ \{\O_i\}~\text{with}~\D_i\lesssim \D_\max\approx 2e^{\g_E/2} \sqrt{c} \sqrt{\log\log c}}
}

\sec{The Bulk}

To summarize, unitary interacting $\cN=2$ SCFTs at large central charge contain an infinite tower of $U(1)_r$-charged, scalar, single-trace, superconformal primary local operators $\{ \O_i\}$ with dimensions $\D_i < \D_*$, where 
\e{Dbound}{\D_* \sim c^{1-{1/ z}} \ll c\,,\qquad 1 \leq z \leq 2}
with $z$ defined by the asymptotic scaling $c\sim r^z$ at large rank $r$ of the Coulomb branch. The number of these operators scales as $\sim c^{1/z}$. The absolute bound for all $\cN=2$ SCFTs (obeying the assumptions of \c{Argyres:2018urp,Caorsi:2018zsq}) is in \eqr{absbound}. These are special (protected) operators, namely, the generators of the Coulomb branch chiral ring, of vanishing $SU(2)_R$ charge and $U(1)_r$ charge $r_i = \D_i$. We have emphasized in \eqr{Dbound} the sub-linear growth of $\D_*$ in the central charge. 

Translating to the semiclassical bulk and taking $a \approx c \sim 1/G_N$, the inverse five-dimensional Newton's constant, \eqr{Dbound} is a bound on the degree of scale separation possible in AdS$_5$ supergravity with at least half-maximal supersymmetry: {\it there must exist an infinite tower of $U(1)$-charged scalar states $\{\phi_i\}$ with energies parametrically below the large black hole threshold.} The latter is set by the mass of large AdS$_5$-Schwarzchild black holes, 
\e{}{E_{\rm BH} \sim \O(1/G_N)\,.}
In particular, all $\phi_i$ obey $E_i < E_*$ where 
\e{zboundgrav}{E_* \sim (E_\BH)^{1-{1/ z}} \ll E_\BH\,.}
It is perhaps clearer to discuss the absolute bound \eqr{absbound} for all such theories, which is (up to a doubly-logarithmic prefactor) the $z=2$ specialization of the above: \textit{there must exist $\sim \sqrt{E_\BH}$ $U(1)$-charged scalar states $\{\phi_i\}$ with energies $E_i < E_* \sim \sqrt{E_\BH}$.}

We emphasize once again that this furnishes a bound on AdS scale separation because it implies the existence of an {\it infinity} of such excitations, a total number that scales with the Planck scale, in the semiclassical limit of weak gravitational coupling. Note that a Planckian number of states in an $\O(1)$ window around Planckian energies is compatible with the existence of the large $c$ limit because the states are not fixed-energy states.\foot{For example, at higher energies $\D \gg E_\BH$, black hole states obey $\N(\D,\D+\d) \sim e^{S(\D)}$ for $\d\sim \O(1)$.} 

\sssec*{Comments}

What are these $U(1)_r$-charged bulk states? In canonical AdS/CFT dual pairs, of course, Coulomb branch generators are dual to KK modes on a large internal space, with energies of order one in AdS units; should the states $\{\phi_i\}$ be realized geometrically, the characteristic scale of extra dimensions is bounded below as
\e{kk}{L_{\rm KK} \gtrsim \ell_p^{3/2}}
where $\ell_p$ is the five-dimensional Planck scale. But remaining steadfastly agnostic in the bootstrap spirit, there are many other possibilities, particularly if the states $\{\phi_i\}$ have Planckian energies. In known AdS$_5$ compactifications of string/M-theory, the spectrum also contains small black holes, with horizon radii obeying $\ell_p \lesssim r_h \ll L_{\rm AdS}$; but non-singular, non-hairy BPS black holes in AdS$_5$ \c{Gutowski:2004ez} must carry large angular momenta, in contrast to the scalar states we are considering. Small black holes are far from the only possibility. In string/M-theory, there is a rich spectrum of semiclassical bulk configurations with energies $M_p \lesssim E \lesssim E_\BH$, including D-branes, giant gravitons \c{Balasubramanian:2001nh}, small black holes (possibly with hair \c{Basu:2010uz,Bhattacharyya:2010yg,Markeviciute:2016ivy,Choi:2023znd}), microstate geometries \c{Bena:2013dka}, topological stars \c{Bah:2020ogh}, grey galaxies \c{Kim:2023sig}, and perhaps yet-undiscovered configurations of matter and black holes in thermal equlibrium. Depending on the energies $\{E_i\}$, some of these may be candidate bulk descriptions of $\{\phi_i\}$. 

We wish to emphasize the conceptual point that BPS states are effective tracers of extra dimensions in AdS/CFT. For one, the HPPS higher-spin gap condition \c{Heemskerk:2009pn} for emergence of AdS$_{d+1}$ Einstein gravity from CFT is blind to the number of large dimensions $D \geq d+1$ of the bulk; but BPS sectors can be sensitive to $D$, because towers of R-charged light states may (and, perhaps, must) be geometrized. More generally, in a $D$-dimensional bulk gravitational effective field theory, the Planckian states are dual to CFT$_d$ operators with conformal dimensions
\e{}{\D \sim M_{p,D} \sim c^{\,{1/(D-2)}}\,.}
These operators are generically unprotected. For example, we recall the short string states in AdS$_5 \x S^5$ at fixed $g_s$, dual to Konishi-type operators in $\cN=4$ SYM with $\D \sim \l^{1/4} \sim c^{1/8}$. So although the $c$-scaling of these unprotected operator dimensions encodes the macroscopic dimensionality of the bulk, it is generally easier to access protected states than unprotected states in CFT. 

Note that \eqr{zboundgrav} is broadly applicable to semiclassical gravity: it is not specific to Einstein gravity, instead relying only on asymptotic scalings valid for any SCFT with ${a\o c} > {1\o2}$. In particular, semiclassical string effects cannot push the compactification scale below \eqr{kk}. In addition, the result applies to SCFTs irrespective of whether they have an independent higher-spin gap scale, $\D_{\rm gap}$ (i.e. it applies to both string and M-theory duals). This generality follows from a mutual compatibility of the known moduli-independence of both the central charges \c{Anselmi:1997am} and the Coulomb branch operator dimensions, built into the $\cN=2$ central charge formulae. 

It is interesting from the gravity point of view that the Coulomb branch dimensions completely fix the strength of the gravitational interaction and the $R^2$ correction (and hence the species scale). This suggests a deeper relation between the Coulomb branch data and BPS black hole sectors that would be nice to understand.

\ssec{Pure supergravity}

Our result also addresses the question of whether AdS$_5$ {\it pure} supergravity exists. It is important to examine what ``pure (super)gravity'' in more than three bulk dimensions ought to even mean. We give an extended discussion of this topic in Appendix \ref{app}.

Low-energy effective field theory provides a natural and agnostic parametric definition of semiclassical AdS$_{d+1}$ pure gravity: namely, the absence of non-graviton states below Planckian energies $E_* \sim M_p$, where $M_p$ is the $(d+1)$-dimensional Planck scale. Although Planckian degrees of freedom lack a universal abstract characterization, the validity of gravitational effective field theory suggests this threshold. Unlike AdS$_3$ gravity, this threshold cannot be made sharp with an $\O(1)$ coefficient. 

Taking the effective field theory view and applying the holographic relation $c \sim M_p^{d-1}$ to the $d=4$ case, a putative CFT dual to AdS$_5$ pure (super)gravity has a (super)conformal primary spectrum comprised solely of (super-)stress tensor composites up to 
\e{}{\D_* \sim c^{1/3}\,.}
Ruling out AdS$_5$ (super)gravity thus amounts to proving that a dual CFT must have a non-stress tensor single-trace primary parametrically below this threshold. We return now to our result \eqr{zboundgrav}. For the range $z<{3\o2}$, we have ruled out AdS$_5$ pure supergravity as the asymptotic limit, violating the gap condition by an infinite tower of sub-Planckian states. Stated conversely, a putative CFT dual to AdS$_5$ pure supergravity must have central charge scaling faster than $c\sim r^{3/2}$ as $r\rar\i$.\foot{The threshold $c \sim r^{3/2}$ is also notable from a different perspective: it is the scaling for which $\delta c_r := c_{r+1} - c_r \sim M_p$. The possible significance of this comment will be motivated elsewhere \c{kitptalk}.} 

There is a sensible analogy to be drawn here with the modular bootstrap in 2d CFTs at large central charge \c{Hellerman:2009bu}. In unitary 2d CFTs, the state-of-the-art upper bound on the first non-vacuum Virasoro primary dimension at $c\rar\i$ is \c{Afkhami-Jeddi:2019zci}
\e{modboot}{\D_* \lesssim {c\o 9.1}\,.}
This sits between the classical threshold at $\D \approx {c\o 12}$ for small BTZ black holes and the onset of universal Cardy thermodynamics at $\D \approx {c\o 6}$ \c{Hartman:2014oaa}. Similarly, our bound \eqr{absbound} for 4d $\cN=2$ SCFTs sits between the Planck scale threshold at $\D\sim c^{1/3}$ for small black holes and the onset of universal AdS$_5$-Schwarzchild thermodynamics at $\D\sim c$. This comparison (meant only to guide the mind) cannot be made apples-to-apples because of the aforementioned differences in dimensional analysis. Relatedly, we have chosen to phrase the CFT$_2$ result more optimistically, in terms of thermodynamics instead of horizon size, because the continuum of geometrically large BTZ black holes, which have horizon radius $r_h \gtrsim \O(1)$ in AdS units, have $\D \approx {c\o 12}(1+r_h^2)$, which is well below \eqr{modboot}. But the bounds are similarly situated with respect to bulk AdS black hole spectra. We expect both bounds \eqr{absbound} and \eqr{modboot} to be sub-optimal. 

Indeed, there are early indications of much stronger bounds on Coulomb branch data of $\cN=2$ SCFTs at $c\rar\i$ that could rule out pure supergravity unconditionally \c{ben}. As noted earlier, $r$-tuples are subject to a host of intricate number-theoretic constraints implied by Coulomb branch geometry. Their systematic exploration at $r\rar\i$ is an intriguing and well-defined open problem in $\cN=2$ SCFT. As a first step in that direction, one can use these constraints to rule out pure supergravity for a specific sequence of large rank SCFTs. In particular, consider a sequence of SCFTs of rank  
\e{}{r = {1\o 2}\prod_{i=1}^N (p_i-1)}
with $N\in\mathbb{N}$, where $p_i$ is the $i$'th prime number. These are the ranks for which $\D_\max$ is an available Coulomb branch generator dimension. Then one can show that if such an $r$-tuple contains $\D_r = \D_\max$, it must also contain a dimension which grows slower than any power of $c$ at $c\rar\i$:
\e{primbound}{\D_r = \D_\max \quad \Rightarrow \quad \D_1 \ll c^\varepsilon~~\forall~~\varepsilon>0\,.}
This excludes a pure supergravity limit for this sequence of SCFTs by a wide margin. How representative this is of generic paths out to $c\rar\i$ in the space of $\cN=2$ SCFTs touches on a subtle question in general abstract CFT. Regardless, \eqr{primbound} indicates the potential for a host of hidden Coulomb branch constraints in the generic case, perhaps strong enough to rule out pure supergravity or even any degree of Planckian scale separation.\foot{We expect even more potential for progress if we combine Coulomb and Higgs branch physics. See \c{Kaidi:2022sng} for an approach to classification of $\cN=2$ SCFTs that does so to impressive effect. As noted there, as $r\rar\i$, the degree $d$ of the modular differential equation governing the Higgs branch VOA seems to scale at least linearly with the rank. Finding an emergent ``effective'' approach at large $r$ and large $d$ poses a difficult but interesting challenge.} 

Ultimately, we would like to understand more deeply the physical mechanism underlying our result, in hopes of extending beyond 4d $\cN=2$ SCFTs. Analytic bootstrap bounds with a bearing on the AdS scale separation question may be more accessible than previously thought.

\sec*{Acknowledgments}

We thank Mario Martone, Leonardo Rastelli and Ben Smith for helpful discussions and related collaboration, as well as the participants of the KITP conference ``\href{https://www.kitp.ucsb.edu/activities/bootstrap-c23}{Gravity from algebra: modern field theory methods for holography}'' where this work was first presented. This research was supported by ERC Starting Grant 853507, and in part by the National Science Foundation under Grant No. NSF PHY-1748958.

\begin{appendix}
\sec{The Appendix}\label{app}

We give a short discussion of some issues in defining ``AdS$_{d+1}$ pure (super)gravity'' in $d>2$. 

In the case of AdS$_3$ gravity, BTZ black holes exist universally above a threshold $\D_* \approx {c-1\o 12}$. In a putative theory of pure gravity, this result is believed to receive corrections non-perturbatively in $1/c$ \c{Maxfield:2020ale}, but is valid to all orders in the semiclassical $1/c$ expansion.\foot{One way to see this is to note that gravitational higher-derivative corrections to the three-dimensional Einstein-Hilbert action can be removed by field redefinitions, to any order in perturbation theory. This implies that the black hole threshold is invariant in a derivative expansion upon using the suitably renormalized relation between $G_N$ and $c$ \c{Kraus:2005vz}. The ``-1'' accounts for graviton loops.} In contrast, there is no universal sharp transition in higher dimensional gravity. The essential difference is seen by dimensional analysis: since $E_\BH \sim \O(M_p^{d-1})$ in AdS$_{d+1}$, there is a parametric gap between the Planck scale and the large black hole threshold when $d>2$. 

In the main text we recalled some of the myriad objects living at energy scales $M_p \lesssim E \lesssim E_\BH$ in string/M-theory. The physics at these scales is non-universal for both IR and UV reasons:

\begin{itemize}

\item Some configurations (e.g. small and/or hairy black holes, other admixtures in thermal equilibrium) depend on the low-energy matter content and/or compactification details. 

\item Others (e.g. D-branes, giant gravitons, extra-dimensional constructions of microstate geometries) encode details of the UV completion. But so far, as we lack a proof of string universality, none of these is known to be required in a UV completion of Einstein (super)gravity. 

\item Finally, there are states that we believe to exist, about which little is known microscopically (e.g. the microscopic description of Planckian states in AdS compactifications of M-theory, dual to maximally supersymmetric CFTs).

\end{itemize}

\ni For these reasons, $E_\BH$ is the threshold for {\it universal} high-energy states in semiclassical Einstein gravity. This is a sense in which the bound \eqr{zboundgrav} is notable. Still, effective field theory suggests $M_p$ as a natural threshold for non-graviton states in ``AdS$_{d+1}$ pure gravity,'' despite leaving some room for interpretation. 

Another feature of the higher-dimensional case relative to AdS$_3$ is that it is even less clear how to define ``pure gravity'' in the {\it quantum} regime, of finite $G_N$. The topological aspect of AdS$_3$ gravity allows for exact computations in certain cases (e.g. chiral gravity partition functions \c{Eberhardt:2022wlc}, or on-shell partition functions of fixed topology saddles in Einstein gravity \c{Collier:2023fwi}) which one may attempt to leverage into a (partial) definition of, or toy model for, a quantum theory. Such a theory would be holgraphically dual to a CFT at finite central charge $c$ which extremizes bootstrap constraints -- a statement which can in principle be made precise. In higher dimensions, gravity is non-topological, and we know of no analogous simplifications. Moreover, the higher-dimensional theories have more parameters than just the central charge $c$. In $d=4$, say, what values should $a/c$ be ``allowed'' to take in a CFT dual to pure {\it quantum} gravity? This point is better phrased with reference to a geometric limit: there is apparent freedom in how $a/c$ is allowed to vary along a sequence of CFTs which approaches pure gravity, and hence $a\approx c$, asymptotically as $c\rar\i$. Similar remarks apply to $t_4$, the coefficient of the third independent tensor structure in the stress tensor three-point function $\< TTT\>$. (In classical Einstein gravity, $t_4 = 0$.) We note that imposing maximal supersymmetry in $d=4$ fixes $a=c$ and $t_4=0$, partially fixing these apparent freedoms, though this simplification fails in $d=6$, in which $a/c$ is not fixed even with maximal supersymmetry (e.g. for the known $A_{N-1}$ (2,0) SCFT, $a=4N^3-3N-1$ and $c={16\o 7} N^3 -{9\o7}N-1$).

\end{appendix}

\end{spacing}

\bibliographystyle{JHEP}
\bibliography{ss_bib}

\end{document}